\documentstyle[11pt,fleqn]{article}
\begin{document}
\title{\LARGE \bf Bootstrap Universe from Self-Referential Noise}  
\author{{Reginald T. Cahill  and     Christopher M. Klinger}\\
  {Department of Physics, Flinders University
\thanks{E-mail: Reg.Cahill@flinders.edu.au, Chris.Klinger@flinders.edu.au}}\\ { GPO Box
2100, Adelaide 5001, Australia }}
\date{ }

\maketitle

\begin{center}
\begin{minipage}{120mm}
\vskip 0.6in
\begin{center}{\bf Abstract}\end{center} {We further deconstruct  Heraclitean Quantum
Systems giving a model for a universe using pregeometric notions  in which
the end-game problem is overcome by means of self-referential noise. The model displays
self-organisation with the emergence of  3-space and time. The time phenomenon is richer
than the present geometric modelling.   }

\vspace{10mm}

\end{minipage} \end{center}

\vspace{50mm}
\hspace{10mm} {\em Ta panta rei - all is flux}
\vspace{3mm}

\hspace{10mm}  Heraclitus of Ephesus, sixth century BC

\newpage

\noindent  {\bf 1. Heraclitean Quantum Systems}  \hspace{5mm}

From the beginning of theoretical physics in the 6th and 5th centuries BC there has been
competition between two classes of modelling of reality: one class has reality   explained in
terms of things, and the other has reality   explained purely in terms of 
relationships (information).  While in conventional physics a mix of these which strongly favours
 the `things' approach
is currently and very efficaciously used,  here we address the problem of the `ultimate'
modelling of reality. This we term the {\it end-game} problem: at higher levels in the
phenomenology of reality one chooses economical and effective models - which usually have to be
accompanied by   meta-rules for interpretation, but at the lower levels   we  are confronted by
the problem of the source of `things' and their rules or `laws'. At one extreme we could have an
infinite regress of ever different `things', another is the notion of a Platonic world where
mathematical things and their rules reside
\cite{Penrose}. In both instances  we still have the fundamental problem of why the universe
`ticks'- that is, why it is more than a mathematical construct; why  is it  experienced? 

 This
`end-game' problem is often thought of as the unification of our most successful and  deepest,
but incompatible,  phenomenologies: General Relativity and Quantum Theory. We believe that the
failure to find a common underpinning of these models is that it is apparently often thought
it would be some amalgamation of the two, and not something vastly different. Another
difficulty is that the lesson from  these models is often confused; for instance from the
success of the geometrical modelling of space and time  it is often argued that the universe
``{\it is} a 4-dimensional manifold''. However the geometrical modelling of time is actually
deficient: it lacks much of the experienced nature of time - for it fails to model both the
directionality of time and the phenomenon of the (local) `present moment'.  Indeed the
geometrical model  might better be thought of as a `historical model' of time, because in
histories the notion of direction and present moment are absent - they must be provided by
external meta-rules. General relativity  then is about possible
histories of the universe, and in this it is both useful and successful.  Similarly quantum
field theories have  fields built upon a possible (historical) spacetime, and subjected to 
quantisation.  But such quantum theories have difficulties with classicalisation and the
individuality of events - as in the `measurement problem'. At best the theory invokes ensemble
measurement postulates as external meta-rules. So our present-day quantum theories are also
historical models. 

 The problem of unifying general relativity and quantum theories then comes
down to going beyond {\it historical} modelling, which in simple terms means finding a better
model of time.  The historical or {\it being} model of reality has been with us since
Parmenides and Zeno, and is known as the Eleatic model. The {\it becoming} or {\it processing}
model of  reality dates back further to Heraclitus of Ephesus (540-480 BC) who argued that 
common sense is mistaken in thinking that the world consists of stable things; rather the
world is in a state of flux. The appearances of `things' depend upon this flux for their
continuity and identity.  What needs to be explained, Heraclitus argued, is not change, but
the appearance of stability.

 Although  `process' modelling  can be traced through to the
present time it has  always been a speculative notion because it has never been implemented in
a mathematical form and subjected to comparison with reality. Various proposals of a {\it pregeometric}
nature have been considered \cite{Wheeler80, Gibbs, Finkelstein}. Here we propose a mathematical
{\it pregeometric process} model of reality - which in
\cite{CK} was called a {\it Heraclitean Quantum System} (HQS).  There we arrived at a HQS by
deconstruction  of the functional integral formulation of quantum field theories  retaining only
those structures which we felt would not be emergent. In this we still started with
`things', namely a Grassmann algebra, and ended with the need to decompose the mathematical
structures into possible histories - each corresponding to a different possible decoherent
classical sequencing. However at that level of the HQS we cannot expect anything other than the
usual historical modelling of time along with its deficiencies. The problem there was that the
deconstruction began with ensembled quantum field theory, and we can never recover individuality
and actuality from ensembles - that has been the problem with quantum theory since its inception. 

 Here we carry the
deconstruction one step further by exploiting  the fact that functional integrals can be
thought of as arising as ensemble averages of Wiener processes. These are normally associated 
with Brownian-type motions in which random processes are used in modelling many-body dynamical
systems.  We argue that  random processes are a fundamental and necessary aspect of
reality - that they arise in the resolution presented here to the end-game problem of
modelling reality. In sect.2 we argue that this `noise' arises as a necessary feature of the
self-referential nature of the universe. In sect.3 we discuss the nature of the self-organised
space and time phenomena that arise, and argue that the time modelling is richer and more
`realistic' than the geometrical model.  In sect.4 we show how the ensemble averaging of
possible universe behaviour is expressible as a functional integral.
\vspace{10mm}

\noindent  {\bf 2. Self-Referential Noise}  \hspace{5mm}

Our proposed solution to the end-game problem is to avoid the notion of things and their rules; rather to
 use a bootstrapped self-referential system. Put simply, this models the universe as a
self-organising and self-referential information system - `information' denoting relationships as
distinct from `things'.   In such a system there is no bottom level and we must  consider the
system as having an iterative character and attempt to pick up the  structure by some
mathematical modelling. 

 Chaitin \cite{Chaitin} developed some insights into the nature of complex self-referential 
information systems: combining Shannon's information theory and Turing's computability theory resulted in the
development  of Algorithmic Information Theory (AIT).  This shows that  number systems contain
randomness and unpredictability, and extends G\"{o}del's discovery, which resulted
from self-referencing problems, of the incompleteness of such systems (see \cite{Zurek90} for
various discussions of the {\it physics of information}; here we are considering {\it information
as physics}).

 Hence if we are
to model the universe as a closed system, and thus  self-referential,
then the mathematical model must necessarily contain randomness. Here we consider one very
simple such model and proceed to show that it produces a dynamical 3-space and a theory for
time that is richer than the historical/geometrical model. 

We model the self-referencing by means of an iterative map
\begin{equation}
B_{ij} \rightarrow B_{ij} -(B+B^{-1})_{ij}\eta + w_{ij},  \mbox{\ \ } i,j=1,2,...,M \rightarrow
\infty.
\label{eq:map}\end{equation} 
We think of $B_{ij}$ as relational information shared by  two monads $i$ and $j$. The monads concept was
introduced by Leibniz, who espoused the {\it relational} mode of thinking in response to and
in contrast to Newton's {\it absolute} space and time.  Leibniz's ideas were very much in the
{\it process} mould of thinking: in this the monad's {\it view} of available  information  and
the commonality of this information is intended to lead to the emergence of space. 
The monad  $i$ acquires its meaning entirely by means of the  information
$B_{i1},B_{i2},...$, where
$B_{ij}=-B_{ji}$ to avoid self-information, and real number valued. The
map in (\ref{eq:map}) has the form of a Wiener process, and the $w_{ij}=-w_{ji}$ are
independent random variables for each $ij$ and for each iteration, and  with variance $2\eta$
for later convenience. The $w_{ij}$   model the self-referential noise.  The beginning of a universe is
modelled by starting the iterative map with $B_{ij} \approx 0$, representing the absence of
information or order. Clearly due to the $B^{-1}$ term iterations will rapidly move the
$B_{ij}$ away from such starting conditions.

The non-noise part of the map involves $B$ and
$B^{-1}$. Without the non-linear inverse term the map would produce independent and trivial random
walks for each
$B_{ij}$ - the inverse introduces a linking of all information. We have chosen $B^{-1}$
because of its indirect connection with quantum field theory (see sec.4) and because of its
self-organising property. It is the conjunction of the noise and non-noise terms which leads to
the emergence of self-organisation: without the noise the map converges (and this determines
the signs in (\ref{eq:map})),  in a deterministic manner to a degenerate condensate type
structure, discussed in
\cite{CK}, corresponding to a pairing of linear combinations of monads.  Hence the map models
a non-local and noisy information system  from which we extract spatial and time-like
behaviour, but we expect residual non-local and random processes characteristic of quantum
phenomena including EPR/Aspect type effects. While the map already models some time-like
behaviour,  it is in the nature of  a bootstrap system that  we start with {\it process}.  In
this system the noise corresponds to the Heraclitean flux  which he also called the ``cosmic
fire'', and from which the emergence of stable structures should be understood.  To Heraclitus 
the flame represented one of the earliest  examples of the interplay of order and disorder. The
contingency and self-ordering of the  process clearly  suggested a model for reality.

\vspace{10mm}

\noindent  {\bf 3. Emergent Space and Time}  \hspace{5mm}

Here we show that the HQS iterative map naturally results in dynamical 3-dimensional
spatial structures.  Under the mapping the noise term will produce rare large
value $B_{ij}$. Because the order term is generally much smaller, for small $\eta$, than the
disorder term these values will persist under the mapping  through more iterations than
smaller valued $B_{ij}$. Hence the larger $B_{ij}$ correspond to some temporary background
structure which we now identify.  

Consider this relational information  from the point of
view of one monad, call it monad $i$.  Monad $i$ is connected via these large $B_{ij}$ to a
number of other monads, and the whole set forms a tree-graph relationship. This is because 
the large links are very improbable, and a tree-graph relationship is much more probable
than a similar graph  with additional links. The simplest 
distance measure for any two nodes within a graph is  the smallest
number of links  connecting  them. Let $D_1, D_2,...,D_L$ be the number of nodes of distance
$1,2,....,L$ from node $i$ (define $D_0=1$ for convenience), where $L$ is the largest distance
from $i$ 
in a particular tree-graph, and let $N$ be the total number of nodes in the tree. Then
$\sum_{k=0}^LD_k=N$. See Fig.1 for an example.

\begin{center}
\begin{picture}(10,180)(+200,50)  
\thicklines

\put(155,205){\line(3,-5){60}}
\put(155,205){\line(-3,-5){60}}
\put(115,140){\line(3,-5){42}}
\put(195,140){\line(-3,-5){21}}

\put(135,200){\Large \bf $i$}
\put(225,200){\Large \bf $D_0\equiv 1$}
\put(225,140){\Large \bf $D_1=2$}
\put(225,100){\Large \bf $D_2=4$}
\put(225,65){\Large \bf $D_3=1$}

\put(155,205){\circle*{10}}

\put(115,140){\circle*{10}}
\put(195,140){\circle*{10}}

\put(95,105){\circle*{10}}
\put(135,105){\circle*{10}}
\put(175,105){\circle*{10}}
\put(215,105){\circle*{10}}

\put(155,70){\circle*{10}}

\end{picture}
\vspace{8mm}

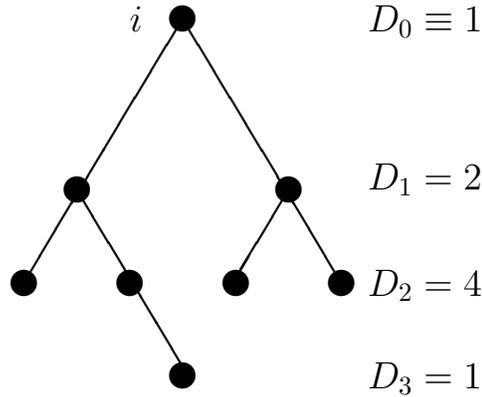
Fig.1 An $N=8$, $L=3$ tree, with indicated distance distributions from  monad $i$.
\end{center}
\vspace{5mm}

 Now consider
the number of different $N$-node trees, with the same distance distribution $\{D_k\}$, to which $i$
can belong.  By counting the different linkage patterns, together with
permutations of the monads we obtain
\begin{equation}
{\cal N}(D,N)=\frac{(M-1)!D_1^{D_2}D_2^{D_3}...D_{L-1}^{D_L}}{(M-N-2)!D_1!D_2!...D_L!},
\label{eq:tcount}\end{equation}
 Here $D_k^{D_{k+1}}$ is the number of different possible linkage patterns between level $k$
and level $k+1$, and $(M-1)!/(M-N-2)!$ is the number of different possible choices for the
monads, with
$i$ fixed. 
 The denominator accounts for those permutations  which have already
been accounted for by the $D_k^{D_{k+1}}$ factors. Nagels
\cite{Nagels} analysed
${\cal N}(D,N)$, and the results imply  that the most
likely tree-graph structure to which a monad can belong  has the distance distribution
\begin{equation}
D_k \approx \frac{L^2\mbox{ln }L}{2\pi^2}\mbox{sin}^2(\frac{\pi k}{L})\mbox{\ \ } k=1,2,...,L.
\label{eq:prob}\end{equation}
for a given arbitrary $L$ value. The remarkable property of this most probable distribution 
is that the $sin^2$   indicates that the tree-graph is embeddable in
a 3-dimensional hypersphere, $S^3$. Most importantly,  monad $i$ `sees' its surroundings
as  being 3-dimensional, since $D_k\sim k^2$ for small $\pi k/L$.  We call these 3-spaces
{\it gebits} (geometrical bits). We note that the $\mbox{ln}L$ factor  indicates that 
larger gebits have a larger number density of points. 

 Now the monads for
which the
$B_{ij}$ are large thus form  disconnected
 gebits.  These gebits however are in turn linked by smaller and more transient $B_{kl}$,
and so on, until at some low level the remaining $B_{mn}$ are  noise only; that
is they will not survive an iteration. Under iterations of the map this spatial network 
undergoes growth and decay at all levels, but with the higher levels (larger $\{B_{ij}\}$
gebits) showing most persistence.  By a similarity transformation we can arrange the gebits
into block diagonal matrices $b_1,b_2,...$, within $B$, and embedded amongst the smaller and
more common noise entries. Now each gebit  matrix has $\mbox{det}(b) =0$, since a tree-graph
connectivity matrix is degenerate. Hence  
 under the mapping the $B^{-1}$ order term has an
interesting dynamical effect upon the gebits since, in the absence of the noise, $B^{-1}$
would be singular. The outcome from the iterations  is that the gebits are seen
to compete and to undergo mutations, for example by adding extra monads to the gebit.
Numerical studies reveal gebits competing  and  `consuming' noise, in a Darwinian process. 

 Hence
in combination the order and disorder terms synthesise an evolving dynamical 3-space with
hierarchical structures, possibly even being fractal.  This emergent 3-space is entirely
relational; it does not arise within any {\it a priori} geometrical background structure.  By 
construction it is the most robust structure,  - however other softer emergent modes of behaviour
will be seen as attached to  or embedded in this flickering 3-space.  The possible fractal
character could be exploited by taking a higher level view: identifying each gebit
$\rightarrow I$ as a higher level monad, with  appropriate informational connections
${\cal B}_{IJ}$, we could obtain a higher level  iterative map of the form (\ref{eq:map}),
with  new order/disorder terms. This would serve to emphasise the notion that in
self-referential systems there are no `things', but rather a complex network of
iterative relations.

In the model the  iterations of the map have the appearance of a cosmic time. However the analysis
to  reveal the internal experiential time phenomenon  is non-trivial, and one would certainly
hope to recover the local nature of experiential time as confirmed by special and
general relativity experiments.   However it is important to notice that the modelling of the time
phenomenon here is much richer than that of the historical/geometric model. First the map is
clearly  uni-directional  (there is an `arrow of time') as there is no way to even define an
inverse mapping because of the role of the noise term, and this is very unlike the conventional
differential equations of traditional physics. In the analysis of the gebits we noted that they
show strong persistence, and in that sense the mapping shows a natural partial-memory phenomenon,
but the far `future' detailed structure of
 even this spatial network is completely unknowable without performing the iterations.
Furthermore the sequencing of the spatial and other structures is individualistic  in that a
re-run of the model will always produce a different outcome. Most important of all is that
we also obtain a modelling of the `present moment' effect, for  the outcome
of the next iteration is   contingent on the noise. So the system shows
overall a sense of a recordable past, an unknowable future and a contingent present moment. 

The HQS process model is  expected to be capable of a better modelling of our experienced reality,
and the key to this is    the noisy processing    the model requires. As well we need the 
`internal view', rather than the `external view' of conventional modelling in physics.
Nevertheless  we would expect that the internally recordable history could be indexed by the
usual real-number/geometrical time coordinate. 

This new self-referential process modelling requires a new mode of analysis  since one cannot
use externally imposed meta-rules or interpretations, rather, the internal experiential phenomena
and the characterisation of the simpler ones by emergent `laws' of physics must be  carefully
determined.  There has indeed been an ongoing study  of how  (unspecified) closed
self-referential  noisy information systems acquire self-knowledge and how the emergent
hierarchical structures can `recognise' the same `individuals' 
\cite{BK}. These {\it Combinatoric Hierarchy} (CH) studies   use the fact that only recursive
constructions  are possible in Heraclitean/Leibnizian systems. We believe that our HQS process
model may provide an explicit representation for the CH studies.      

\vspace{10mm}
\noindent  {\bf 4. Possible-Histories Ensemble}  \hspace{5mm}

While the actual history of the noisy map can only be found in a particular `run', we can
nevertheless show that  averages over an ensemble of possible histories can be determined,
and these have the form of functional integrals.  The notion of an ensemble average 
for any function $f$ of the $B$, at iteration $c=1,2,3,...$, is expressed by
\begin{equation}
<f[B]>_c=\int {\cal D}Bf[B]\Phi_c[B],
\label{eq:enav}\end{equation}
where $\Phi_c[B]$ is the ensemble  distribution. By the usual construction 
for Wiener processes we obtain the Fokker-Planck equation 
\begin{equation}
\Phi_{c+1}[B]=\Phi_{c}[B]-\sum_{ij}\eta\left(\frac{\partial}{\partial
B_{ij}}((B+B^{-1})_{ij}\Phi_{c}[B])-\frac{\partial^2}{\partial B_{ij}^2}\Phi_{c}[B]\right).
\label{eq:FP}\end{equation}
For simplicity, in the quasi-stationary regime, we find
\begin{equation}
\Phi[B] \sim \mbox{exp}(-S[B]),
\label{eq:stat}\end{equation}
where the action is
\begin{equation}
S[B]=\sum_{i>j}B_{ij}^2-\mbox{TrLn}(B).
\label{eq:S}\end{equation}
Then the ensemble average is
\begin{equation}
\frac{1}{Z}\int {\cal D}Bf[B]\mbox{exp}(-S[B]),
\label{eq:FIC}\end{equation}
where $Z$  ensures the correct normalisation for the averages. The connection between
(\ref{eq:map}) and (\ref{eq:S}) is given by
\begin{equation}
(B^{-1})_{ij}=\frac{\partial}{\partial B_{ji}}\mbox{TrLn}(B)=
\frac{\partial}{\partial B_{ji}}\mbox{ln}\prod_{\alpha}\lambda_{\alpha}[B].
\label{eq:div}\end{equation}
which probes the sensitivity
of the invariant ensemble information to changes in $B_{ji}$, where the information is in the  
 eigenvalues $\lambda_{\alpha}[B]$ of $B$. A further transformation is possible \cite{CK}:
\begin{eqnarray*}
<f[B]>=\frac{1}{Z}\int{\cal D}\overline{m}{\cal D}m{\cal D}B \mbox{\
}f[B]\mbox{exp}\left(-\sum_{i>j}B_{ij}^2+
\sum_{i,j}B_{ij}(\overline{m}_im_j-\overline{m}_jm_i)\right),
\end{eqnarray*}
\begin{equation}
=\frac{1}{Z}f[\frac{\partial}{\partial J}]\int{\cal D}\overline{m}{\cal D}m\mbox{\ }
\mbox{exp}\left(-\sum_{i>j}\overline{m}_im_j\overline{m}_jm_i+\sum_{ij}J_{ij}(\overline{m}_im_j-
\overline{m}_jm_i)\right).
\label{eq:Grass}\end{equation}
 This expresses 
the ensemble average in terms of an anticommuting Grassmannian algebraic  computation
\cite{CK}. This suggests how the noisy information map may lead to fermionic modes. While
functional integrals of the above forms are common in quantum field theory, it is significant
that in forming the ensemble average we have lost the contingency or present-moment
effect.  This always happens - ensemble averages do not tell us about individuals - and
then the meta-rules and `interpretations' must be supplied in order to generate some notion of
what an individual might have been doing. 

The Wiener iterative map can be thought of as
a resolution of the functional integrals into different possible histories. However this does
not imply the notion that in some sense {\it all} these histories must be realised, rather only
{\it one} is required. Indeed the basic idea of the process modelling is that of individuality. 
Not unexpectedly we note that the modelling in (\ref{eq:map}) must be done from within that {\it
one} closed system.

 In
conventional quantum theory it has been discovered that  the  individuality of the measurement
process - the `click' of the detector - can be modelled by adding a noise term to the Schrodinger
equation
\cite{GP}. Then by performing an ensemble average over many individual runs of this modified
Schrodinger equation one can derive the ensemble measurement postulate - namely
$<A>=(\psi,A\psi)$ for the ``expectation value of the operator A''.  This individualising of
the ensemble average has been shown to also relate to the decoherence functional formalism \cite{Diosi}. 
There are a number of other proposals considering noise in spacetime modelling \cite{Percival,Cal}.

\vspace{10mm}
\noindent  {\bf 5. Conclusion}
 
We have addressed here the unique end-game problem which arises when we attempt to model and comprehend
the universe as a closed system. The outcome is the suggestion that the peculiarities of this end-game
problem are directly relevant to our everyday experience of time and space; particularly the
phenomena of the contingent present moment and the three-dimensionality of  space.  This analysis
is based upon the basic insight that a closed self-referential system is necessarily noisy. This
follows from Algorithmic Information Theory. To explore the implications we have considered a
simple {\it pregeometric  non-linear noisy iterative map}. In this way we construct a process
bootstrap system with  minimal  structure. The analysis shows that the first self-organised
structure to arise  is a dynamical 3-space formed from competing pieces of 3-geometry - the
gebits.  The analysis of experiential time is more difficult, but it will clearly be a contingent
and process  phenomenon which is more complex than the current geometric/historic modelling of
time.  To extract emergent properties of self-referential systems requires that an internal view
be considered, and this itself must be a recursive process. We suggest that the
non-local self-referential noise has been a  major missing component of our modelling of reality. 
Two particular  applications are an understanding of why quantum detectors `click' and  of the
physics of consciousness  \cite{Penrose}, since both  clearly have an essential involvement with
the modelling of the present-moment effect, and cannot be understood using the geometric/historic
modelling of time.

We thank Susan Gunner and Khristos Nizamis for useful comments. Research supported by an ARC Small
Grant from Flinders University.


\newpage

\begin{thebibliography}{99}

\bibitem{Penrose}  R. Penrose, {\it The Large, the Small and the Human Mind}, (Cambridge Univ.
Press 1997).

\bibitem{Wheeler80}  J.A. Wheeler, {\em Pregeometry: Motivations and Prospects},  in {\em
Quantum Theory and Gravitation}, A. R. Marlow, ed. (Academic Press, New York 1980).



\bibitem{Gibbs}  P. Gibbs, {\em  The Small Scale Structure of Space-Time: A bibliographical
Review}, hep-th/9506171.

\bibitem{Finkelstein} D.R. Finkelstein, {\em Quantum Relativity}, (Springer, 1996).

\bibitem{CK} R.T. Cahill and C.M. Klinger, Phys. Lett. A {\bf 223}(1996)313.


\bibitem{Chaitin}  G.J. Chaitin, {\it Information, Randomness and Incompleteness},  2nd ed.
(World Scientific, 1990).

\bibitem{Zurek90} W.H. Zurek (ed), {\it Complexity, Entropy and the Physics of
                    Information}, (Addison-Wesley,  1990).



\bibitem{Nagels} G. Nagels, Gen. Rel. and Grav. {\bf 17}(1985)545.

\bibitem{BK} T. Bastin and C. Kilmister, {\it Combinatorial Physics}, (World Scientific, 1995).

\bibitem{GP}  N. Gisin and I.C. Percival, {\it  Quantum State Diffusion:
From Foundations to Applications}, quant-ph/9701024.

\bibitem{Diosi}  L. Diosi, N. Gisin, J. Halliwell, I.C. Percival,
Phys. Rev. Lett. {\bf 74}(1995)203.  

\bibitem{Percival} I.C. Percival, Proc. R. Soc. Lond. A {\bf 453}(1997)431.

\bibitem{Cal} F. Calogero, Phys. Lett. A {\bf 228}(1997)335.

\end{thebibliography}
\end{document}